\input psfig
\input cp-aa.tex
\def\msol{{\cal M}_\odot}
\def\ueber#1#2{{\setbox0=\hbox{$#1$}%
  \setbox1=\hbox to\wd0{\hss$ #2$\hss}%
  \offinterlineskip
  \vbox{\box1\box0}}{}}
\def\lesssim{\,\lower 1mm \hbox{\ueber{\sim}{<}}\,}
\def\grsim{\,\lower 1mm \hbox{\ueber{\sim}{>}}\,}

%
  \MAINTITLE={X-ray analysis of Abell 2634 and its central galaxy 3C465}
  \AUTHOR={Sabine Schindler@{1,2,*} \& M. Almudena Prieto@{1,**}} 
  \INSTITUTE={ 
@1 Max-Planck-Institut f\"ur extraterrestrische Physik,
   Giessenbachstra\ss e, D-85748 Garching, Germany
@2 Max-Planck-Institut f\"ur Astrophysik, Karl-Schwarzschild-Stra\ss e 1,
   D-85748 Garching, Germany
}
  \catcode`\@=12
  \FOOTNOTE{e-mail: sas@mpa-garching.mpg.de}
  \FOOTNOTE{e-mail: alm@mpe-garching.mpg.de}
  \catcode`\@=11
  \ABSTRACT={An analysis of a ROSAT/PSPC observation of the galaxy
cluster A2634 is presented. It has a luminosity of
$7.9\pm0.1\times10^{43}$erg/s  in the ROSAT band (0.1-2.4 keV). The
temperature profile decreases from about 3 keV in the outer
parts to 1.2 keV in the centre. Within a
radius of 1.5 Mpc the gas mass of the cluster is  0.51$\times
10^{14}\msol$ and the total mass amounts to $4.1^{+2.6}_{-1.8}\times
10^{14}\msol$. The X-ray morphology shows two peculiar features -- a
strongly peaked emission in the centre and an excess emission in the
south-west. The central emission probably originates from a weak cooling flow,
the SW emission is possibly associated with higher density regions in
pressure equilibrium with the intra-cluster medium. This region of
excess emission is somewhat embraced by   
 the wide-angle radio tails associated with 3C465 suggesting
that the overpressure of the
relativistic particles has displaced the thermal gas.}
  \KEYWORDS={Galaxies: clusters: individual: A2634 -
Galaxies: individual: 3C465 -
inter-galactic medium - dark matter 
 -  X-rays: galaxies - Cosmology: observations} 
  \THESAURUS={03 {11.03.4; 11.09.1; 11.09.3; 12.04.1; 13.25.2;  12.03.3} }
%
%
%
\maketitle
%

\titlea{Introduction}

A2634 is a nearby Abell cluster at z=0.0293 (Scodeggio et al. 1995) of  
richness class I. 
 The dominant central object is the radio-galaxy 3C465 (or NGC 7720), 
a prototype
wide-angle tailed
 (WAT) source of moderate radio power, $ F(178 {\rm MHz}) = 6.4\times10^{37}$
Jy (Laing et al. 1983) and narrow optical emission
lines.

A2634 is projected onto a complex region in the sky dominated by  large-scale 
structure associated
with the Pisces-Perseus supercluster.  
A detailed description of the 12 degrees  field around
A2634 can be found in  Scodeggio et al. (1995). Briefly, the region includes 
several clusters and groups; yet, none
of them are found gravitationally bound to A2634.

The spatial distribution of the cluster members, their kinematics and
dynamics were studied extensively in  Scodeggio et al's work on
the basis of optical and radio data. The authors find the properties
of the early-type population component being in agreement with those
expected from a relaxed system, whereas the opposite is found for the
spiral population, whose multimodal velocity distribution and lack of
concentration towards the cluster core suggest their recent arrival to 
the system.  However,  the lack of significant
clumpiness in the galaxy distribution argues against a
merger event to have occurred in the plane of the sky. A remaining
possible option
would be the possibility that loose groups of spirals were falling
 onto the cluster along the line of sight.
 
Pinkney et al. (1993) provided a different view of A2634. The authors
reported on a peculiar radial velocity of more than 200 km/s to be
associated with the central galaxy. They  found substructure in
the outer parts of the cluster and concluded that A2634 is an
unrelaxed cluster with an ongoing merger.
Burns et al. (1993) explained the radio and X-ray appearance of A2634
as being in a post-merger state.

In this paper, we further explore the dynamical state of A2634
on the basis of a detailed analysis of the ROSAT/PSPC  data of the
cluster. The  spatial distribution of the PSPC X-ray light and the
associated dynamics are compared with the spatial distribution of 
the cluster galaxies and their kinematics as derived from the optical study by
  Scodeggio et al. (1995). Furthermore, the PSPC 
spectral information is used to determine the
temperature of the intra-cluster gas and cluster mass. 
Finally, possible origins for the bending tailed radio
structure associated with the central galaxy and its relationship with
the hot cluster medium are discussed.

$H_0=50$ km/s/Mpc is used throughout. 1 arcsecond is about 850 pc at the
distance of A2634.

\titlea{Data}

\begfigwid 0.0cm
\psfig{figure=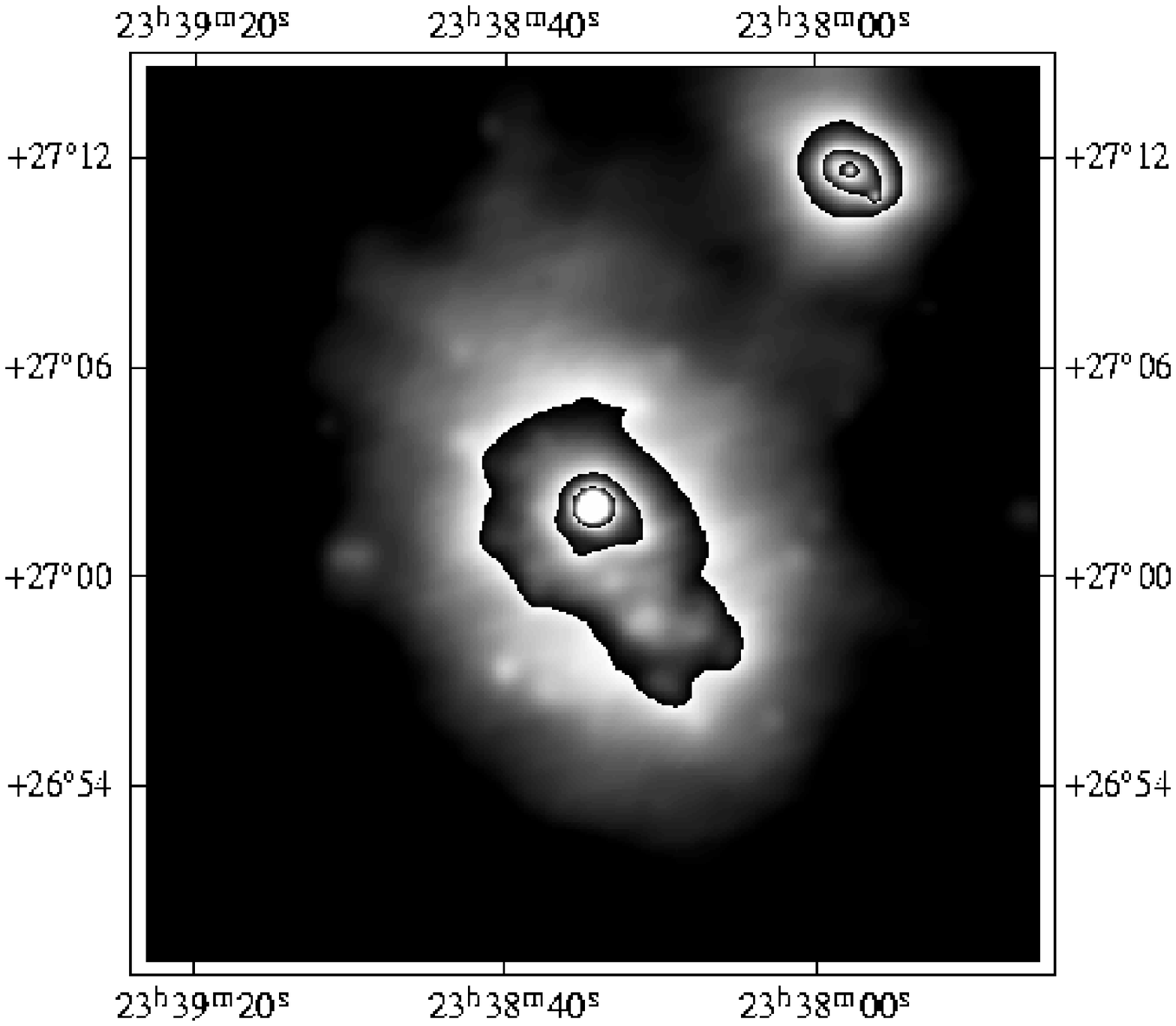,width=17.0cm,clip=} 
\figure{1}{ROSAT/PSPC image of the cluster A2634 in the ROSAT hard
energy band (0.5-2.0 keV).
It is smoothed with a Gaussian filter of 
$\sigma$ varying from 1.8
arcseconds to 1.4 arcminutes. The
X-ray image is dominated by the emission centred on 3C465 and 
shows substructure in the south-west.
The emission in the north-west of A2634 is associated with
the cluster CL-37.
}
\endfig

A2634 was observed in a pointed observation with the ROSAT/PSPC
 on June
20$^{\rm th}$ to 22$^{\rm th}$, 1991. The total effective exposure
time was 9100 sec.  The central galaxy of the cluster, 3C465, was
placed at the centre of the PSPC field-of-view. 
To flat-field the image,
corresponding energy-weighted exposure maps are derived and
subsequently combined according to the energy distribution of the source.
The data are then  divided by these maps.
Finally, a Gaussian filter with 
$\sigma$ varying from 1.8 arcseconds to 1.4 arcminutes
is applied
to the data. The resulting  accumulated image in the 0.5-2 keV
band is shown in Fig. 1.

The most prominent emission feature in the X-ray  image  is at the 
centre and
coincides with the position of 3C465 and its companion galaxy, the latter 
being located $\sim$ 12 arcseconds to the north
from  3C465 (Fig. 2). HST/WFPC2 images of the field taken
in the R band reveal no signs of interaction between the two objects,
both showing well defined elliptical isophotes. The two
galaxies both show prominent rings of dust (Fig 2.). 
VLBI observations (Venturi et al. 1995)
reveal the central conspicuous
morphology of the core region of 3C465. 

The ROSAT/HRI image of the cluster 
(Sakelliou \& Merrifield 1997) also shows one single point source  at the 
centre of the cluster.  We thus expect  the peak of
the central X-ray emission to be coincident with the  3C465 radio
core position.
 
\begfig 0.0cm
\psfig{figure=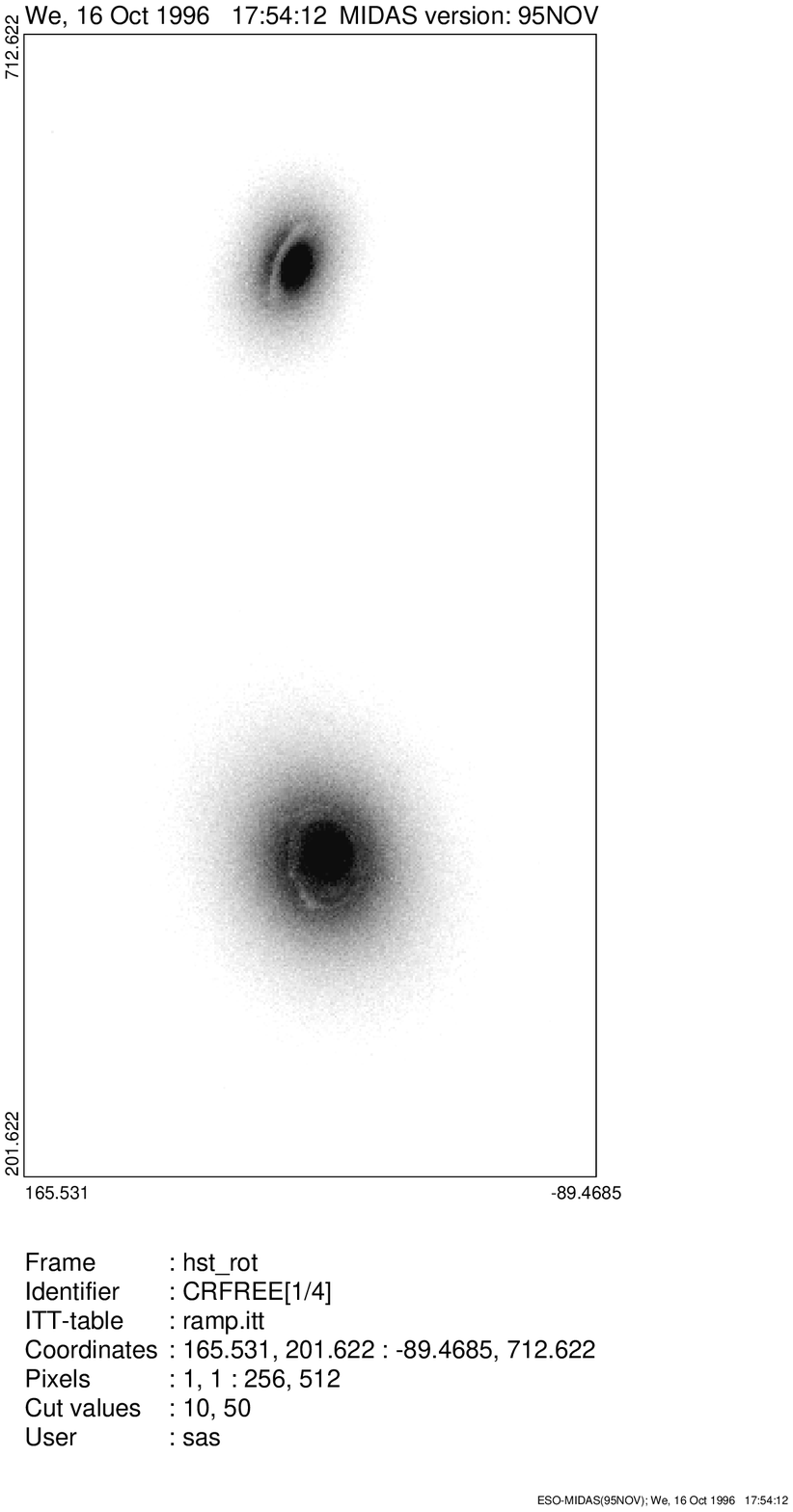,width=6cm,clip=} 
\figure{2}{HST/WFPC2 image of 3C465 (bottom) and its companion
galaxy observed with the F702W filter. Both galaxies show prominent dust lanes.
The distance between the centres of the two galaxies is 12 arcseconds.
}
\endfig

The second bright source in Fig. 1 (upper-right corner of the figure) is
associated with the background cluster CL-37 (z$\sim$0.123, Scodeggio
et al. 1995), which is projected $\sim$ 13 arcminutes from the centre of
A2634.

The diffuse X-ray emission associated with A2634 shows clearly an 
elongation in
NE - SW direction. This asymmetry is particularly pronounced
within the $\sim$ 6 arcminutes inner radius. 
Excess emission above the diffuse average level 
is apparent in the south-west region of the cluster, where
  a somewhat  elongated and patchy
structure is clearly seen (Fig. 1). 
This excess emission was already seen  in previous EINSTEIN
IPC  maps of the cluster (Eilek et al. 1984) though  with
 much less 
spatial detail  due to the poorer spatial 
resolution of the  IPC (FWHM $\sim 1.5$ arcmin).

Fig. 3  presents a logarithmic contour image of the central field shown
in Fig. 1, also from the same energy band, 0.5-2 keV.
The 5 GHz VLA map by Venturi et al. (1995) is overlaid (see also Burns
et al. 1993).  The inferior
resolution of the PSPC hampers any detailed comparison between the
radio and X-ray data; yet, one can readily see that the overall
distribution of the extended radio emission avoids somewhat that of
the X-ray emission. Although projection effects can affect the
comparison, in the apparent distribution the    
kiloparsec-scale jet associated with the
central galaxy 3C465 extends almost perpendicular to the main
direction of the X-ray emission. Both
radio tails give the impression to open their way through the hot medium by
bending and propagating along regions of lower X-ray surface brightness.
In particular, the X-ray surface brightness is low in the direction of
the tails around a radius of about 1 arcminute (see also Fig. 6).

\begfig 0.0cm
\psfig{figure=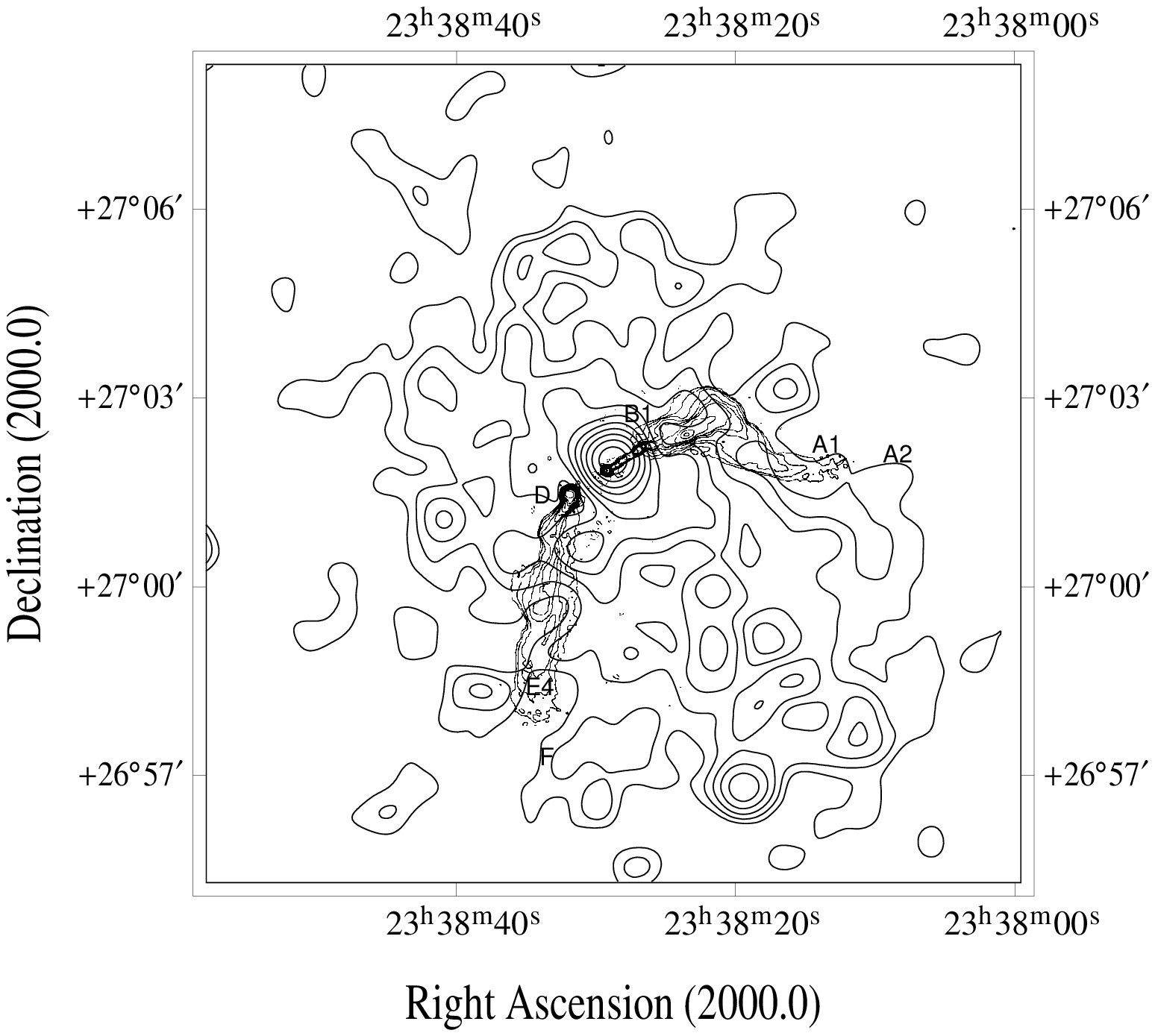,width=8.8cm,clip=} 
\figure{3}{5 GHz VLA map (Venturi et al. 1995) superposed on
contours of a PSPC image smoothed with a Gaussian filter of $\sigma=$
15 arcsec. The spacing of the X-ray
contours is logarithmic with 10 contours per order of magnitude. The
highest contour line corresponds to a countrate of $9.2\times10^{-4}$
counts/s/arcmin$^2$. 
}
\endfig

\titlea{Spatial analysis}

The X-ray emission in A2634
can grossly be resolved in three main components: a symmetric 
 region of diffuse emission
extending into   the Mpc range, an inner region of enhanced emission and
 peculiar morphology distributed 
in south-westerly direction extending up to
about 6 arcminutes ($\sim$ 300 kpc) from the centre, and finally, an
unresolved peaked emission at the centre.
The corresponding ROSAT/HRI image of the cluster 
(cf. Fig. 1 in Sakelliou \& Merrifield 
1997) confirms 
 the peaked central emission and the elongation of the overall 
X-ray structure in the NE-SW direction. 

To quantify the asymmetry of the  
diffuse X-ray emission, 
ellipses of different isophote levels are fitted to the PSPC image
(cf. Bender \&
M\"ollenhoff 1987). Relatively high eccentricities of 0.3 -
0.5 between the surface brightness levels $1.6\times 10^{-3}$ and
$4.3\times 10^{-3}$ cts/s/arcmin$^2$ (Fig. 4) are found, 
which are caused by the excess
emission to the south-west of the maximum. The ellipses within this interval
have position angles between 20$^\circ$ and 50$^\circ$ (north is 0,
counterclockwise). These angles agree well with the outer optical
isophotes of 3C465 (de Juan et al. 1994) 
although the X-ray isophotes reach up to much
larger radii (the major axes in this interval are between 2.3 and 9.6
arcminutes). The agreement between the optical and X-ray rotation angles 
was previously found  from the  EINSTEIN IPC X-ray
image (Pinkney et al. 1993).
 
\begfig 0.0cm
\psfig{figure=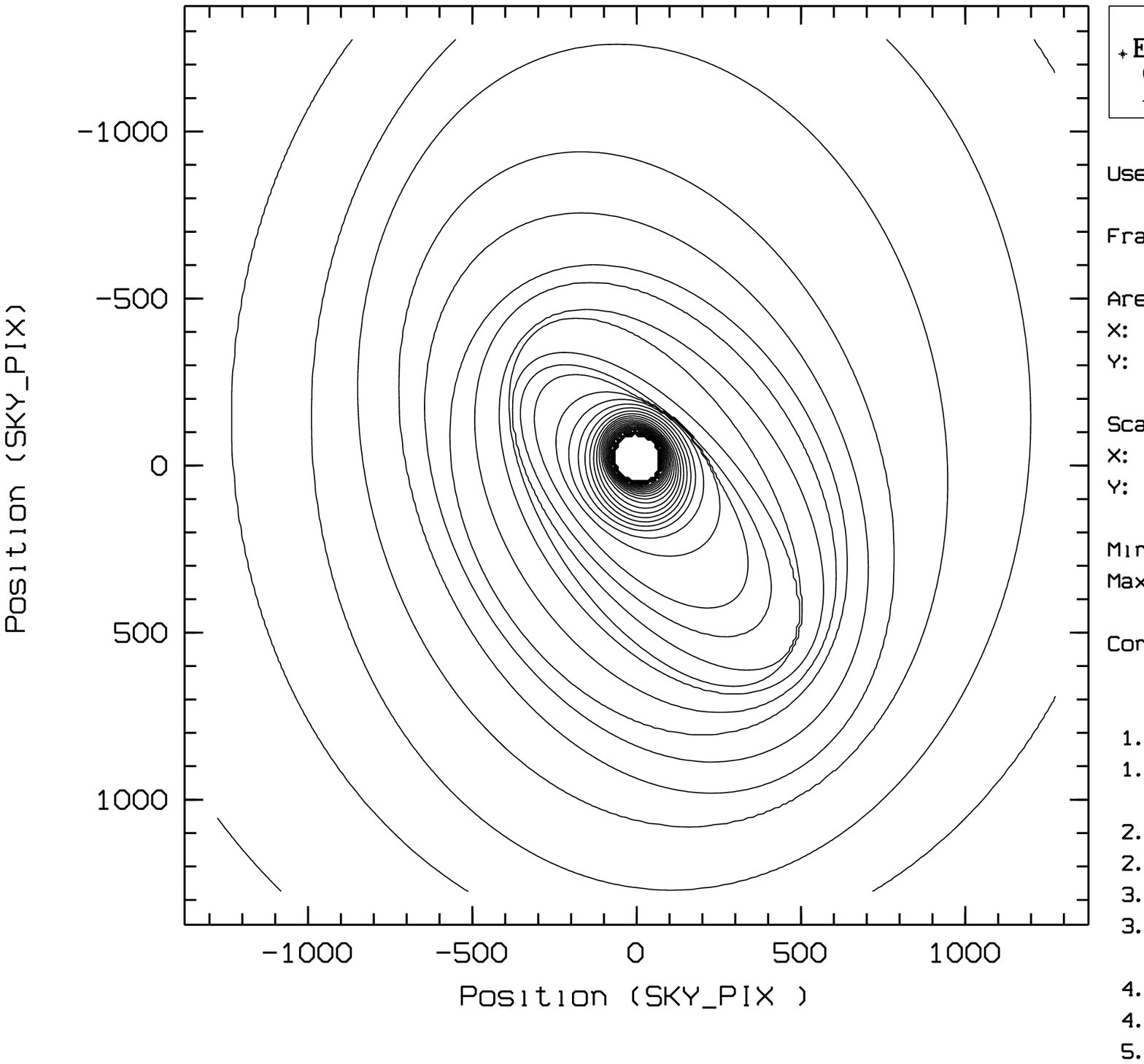,width=8.8cm,clip=} 
\figure{4}{Ellipse fits to different isophote levels. The original
image is exposure corrected and smoothed with a Gaussian filter with
$\sigma = 25$ arcsec. We find large eccentricities of 0.3 - 0.5 in
NE-SW direction at 
surface brightness levels of $1.6\times 10^{-3}$ to
$4.3\times 10^{-3}$ cts/s/arcmin$^2$. The size of the image is $21.3\times21.3$
arcmin$^2$.} 
\endfig

As the outer region  of the cluster  shows  small eccentricities,
the cluster surface brightness profile is derived assuming spherical symmetry 
 (Fig. 5). As centre, the
maximum of the X-ray emission at $\alpha=23^h38^m29^s$, 
$\delta=+27^{\circ}02'00''$ (J2000) is used. It differs by about 8 arcseconds 
from the radio core coordinates  (Venturi et al. 1995), which is
consistent with the expected ROSAT pointing errors. Only  photons in the 
ROSAT hard band (0.5-2.0 keV) are considered. 
Nine sources around the cluster which are probably 
fore- or background
sources are  excluded from the profile extraction. Two of these nine
sources are seen in Fig.1: CL-37 and the point-like source on the central-left 
side of the image at $\alpha=23^h39^m00^s$,  
$\delta=+27^{\circ}00'40''$ (J2000).

The resulting light profile is fitted with a 
$\beta$-model (Cavaliere \& Fusco-Femiano 1976; Jones \& Forman 1984):

$$\Sigma (r) = 
\Sigma_0 \left( 1 + {\left({r \over r_c}\right)}^2\right)^{-3\beta + 1/2},
   \eqno(1)
$$
where $\Sigma_0$ is the central surface brightness, $r_c$ is the core
radius, and $\beta$ is the slope parameter. The central bins ($r<30$
arcseconds) are excluded for the fit because they are obviously
dominated by the emission of the central source. A fit of the region between
30 arcseconds up to 5 arcminutes from the centre yields $\Sigma_0=3.4\times 
10^{-3}$
counts/s/arcmin$^2$, $r_c= 9.8^{+1.3}_{-1.1}$arcmin
($490^{+70}_{-50}$kpc), and $\beta = 0.79^{+0.10}_{-0.07}$ (1$\sigma$
errors). The results for the fit parameters do not change
significantly when varying the radial binning.

\begfig 0.0cm
\psfig{figure=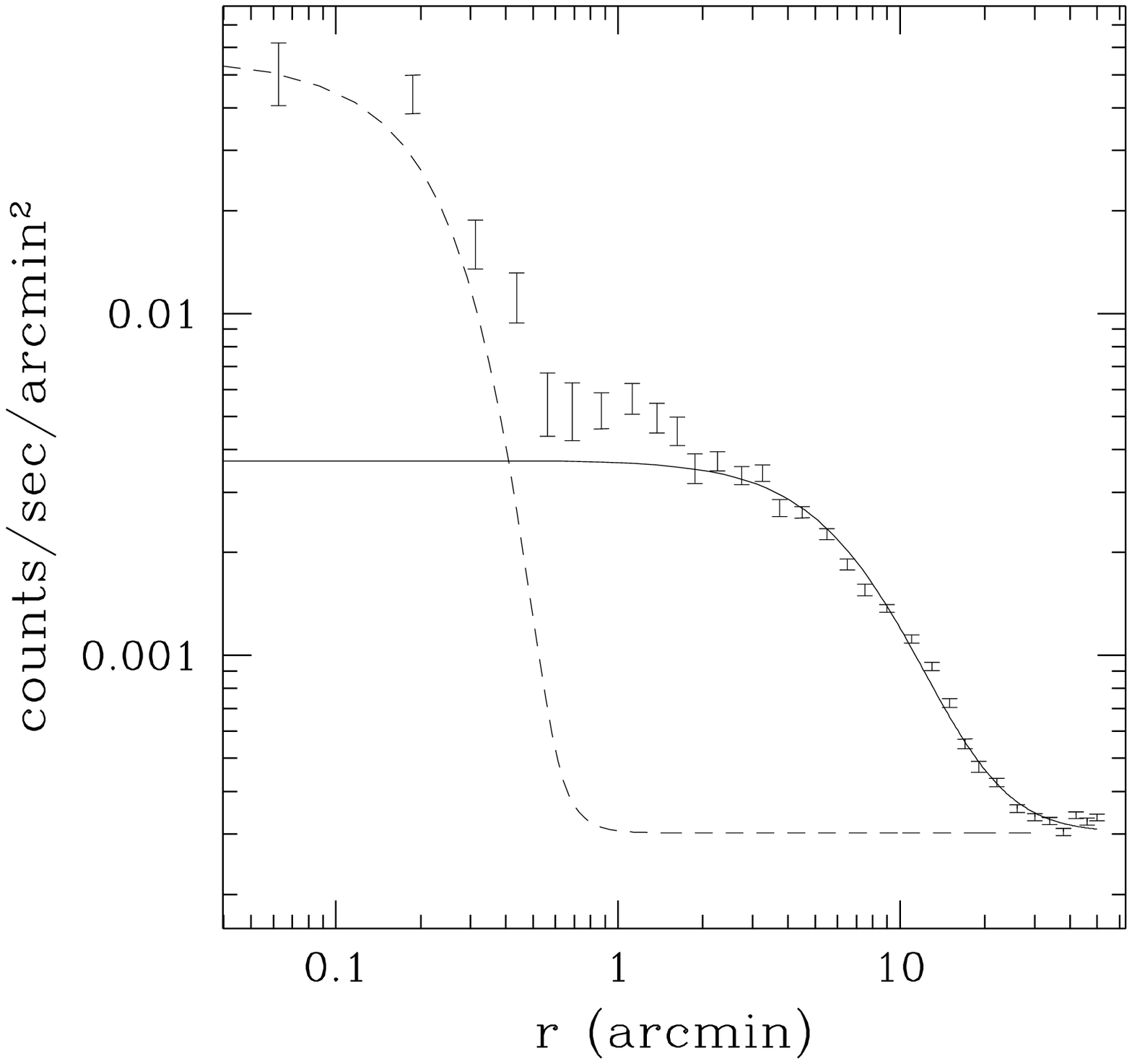,width=8.8cm,clip=} 
\figure{5}{Radial profile of the hard band (0.5-2.0 keV) 
emission of A2634. The
solid line is a fit with a $\beta$-model. The four innermost bins are
ignored by the fit because they are dominated by the central source. 
Owing to the elongation of the cluster 
we find a relatively large core radius of
$r_c=490$kpc. For comparison also the point spread function of the
ROSAT/PSPC normalized to the central bin is plotted (dashed line).
}
\endfig

Taking into account that the cluster is more extended in 
south-westerly direction, 
independent  fits are applied to the 
  south-west and  north-east profiles, respectively. In the
north-east (-45$^\circ$ to 135$^\circ$), very similar values to
the overall profile values are found: $\Sigma_0=3.1\times 10^{-3}$
counts/s/arcmin$^2$, $r_c= 9.7$arcmin, and $\beta = 0.75$, whereas the fit to
 the south-west profile gives a larger core radius as well as a larger 
$\beta$, which  reflects
the observed elongation in this direction. The resulting fit parameters in this case are:
 $\Sigma_0=3.6\times 10^{-3}$
counts/s/arcmin$^2$, $r_c= 11.0$ arcmin, $\beta = 0.91$.

To make the excess emission visible,  a synthetic, spherically
symmetric  image with 
 the parameters of the overall 
$\beta$-fit is constructed.
 When subtracting the synthetic image from the
 PSPC image, two separated regions of enhanced emission are
readily seen (Fig. 6). Besides the unresolved emission in the centre,
enhanced emission along the NE-SW direction is revealed, the excess being more
prominent towards the south-west region of the
cluster. Note that the most prominent regions of  enhanced emission
 are  already apparent in the original PSPC image (Fig. 1 and 3).
The significance of this excess emission is determined using  the method
developed by Neumann \& B\"ohringer (1997). A 
 significance of
10$\sigma$ is derived for the central emission,  the two maxima in the
south-west part having significances of 5$\sigma$ and 4$\sigma$, respectively.
Besides, the derived model  subtracts fairly well the 
 X-ray distribution in other parts of the cluster, indicating that the
diffuse intracluster emission is compatible with spherical symmetry.

\begfig 0.0cm
\psfig{figure=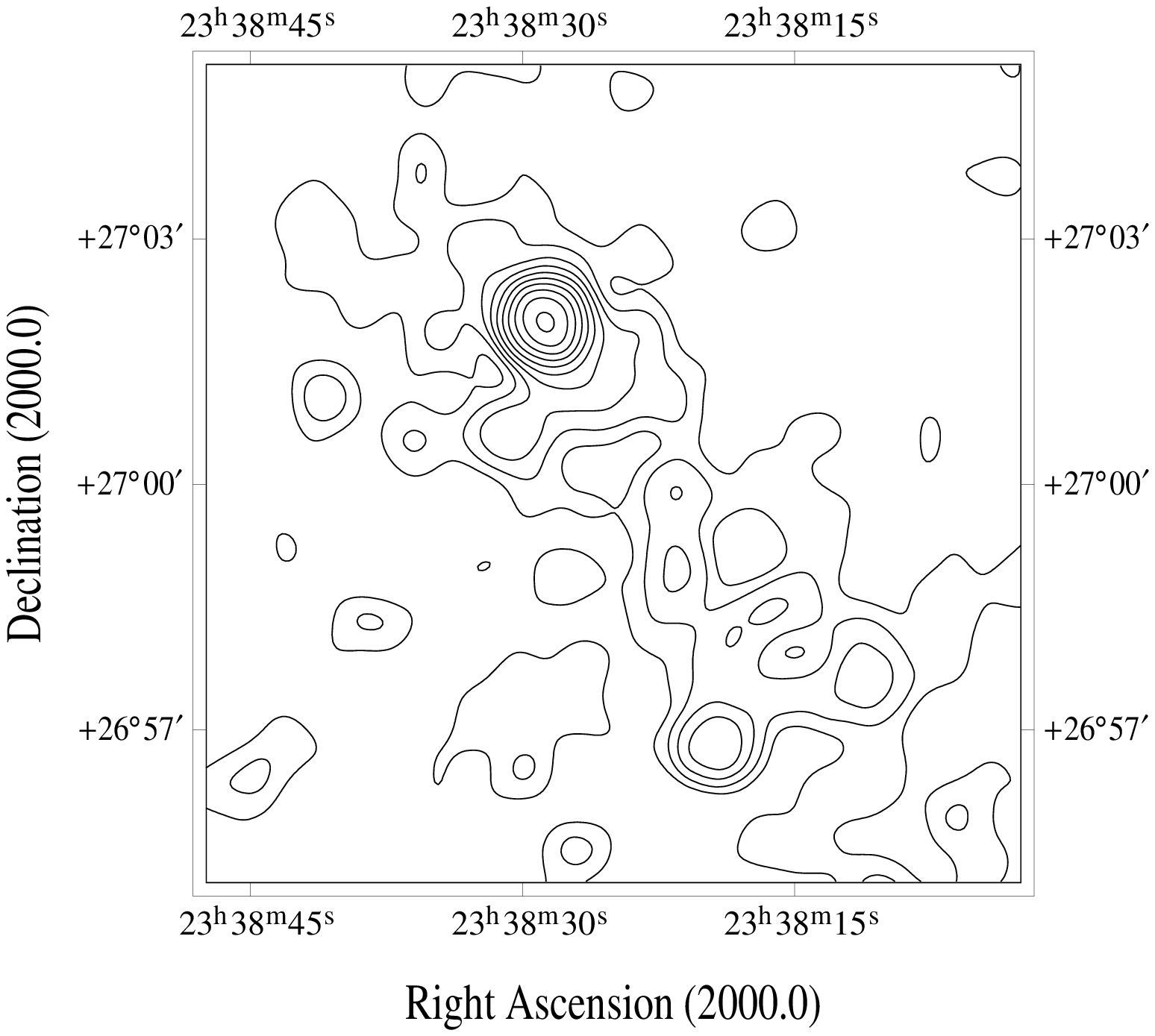,width=8.8cm,clip=} 
\figure{6}{Residual image after subtracting a spherically symmetric
image constructed with a $\beta$ profile
with $\Sigma_0=3.4\times 10^{-3}$
counts/s/arcmin$^2$, $r_c= 9.8$arcmin, and $\beta = 0.79$ centred on
the X-ray maximum from an image smoothed with Gaussian filter
of $\sigma=15$arcsec. 
The contours represent the significance of emission above the
$\beta$ profile in units of $\sigma$.
The excess emission in the centre has a significance of
10$\sigma$. The two maxima in the
south-west have significances of 5$\sigma$ and 4$\sigma,$ respectively.
}
\endfig

\titlea{Spectral analysis}

The brightness and proximity of A2634  permits the collection of
a sufficient number of photons for a spectral analysis to be
performed. The primary objective is to determine
whether a temperature variation does exist across  
the cluster. Individual spectra are accumulated from different
regions in the cluster.
The size and the number of  regions are a
compromise between adequate spatial resolution 
 and the
inclusion of a sufficient number of counts for a reliable fit to be
performed. 

Individual spectra are extracted from
 three consecutive concentric rings: 
an outer one,
comprising the region between 5.5 and 9.5 arcmin, an intermediate one
between 3 and 5.5 arcmin, and an inner one, between 1.5 and 3 arcmin.
The spectrum of the central region is derived from a region of
 1 arcminute radius. In addition, an average temperature of the cluster
is derived from a fit to  
the whole diffuse emission region excluding a central region of
 1.5 arcminutes radius.

Each spectrum is fitted with a Raymond \& Smith model 
(1977). The limited number of photons within each of the rings together
with the relatively soft energy range of the PSPC prevents a
simultaneous fitting of all parameters: metallicity, hydrogen
column density, temperature and the flux normalization  factor.  Thus,
 the metallicity is fixed to a typical cluster value of
$m=0.35$ (Arnaud et al. 1992; Yamashita 1994; Tsuru et al. 1996).
Table 1 summarizes the main fit results.

\begtabfull
\tabcap{1}{PSPC spectral analysis: parameters are derived from a single 
Raymond-Smith model with fixed metallicity $m=0.35$. Errors are 1 
sigma correlated multi-parameters uncertainties. Fit$^a$ corresponds to a 
power-law fit,  the parameter marked is the power-law index. 
SW is the south-west sector defined by
 angles from 180 to 245 degrees (north over east) and radii as indicated. }
\halign{#\quad\hfil&#\quad\hfil&#\quad\hfil&#\hfil&#\hfil \cr
\noalign{\hrule\medskip}\cr
radius & total & temperature& $n_{\rm H}$ & $\chi{^2}/dof$ \cr
[arcmin]&counts & [keV] &  [$10^{21}$cm$^{-2}$] &  \cr
$1^{'}\sim$ 50 kpc &  \cr
\noalign{\hrule\medskip}\cr
1.5 - 11.0& 5700 & $3.5^{+1.2}_{-0.7}$ & $0.6\pm0.1$ & 1.0/90 \cr
5.5 - 9.5 & 2500 & $3.5^{+2.0}_{-1.0}$ & $0.4^{+0.2}_{-0.0}$ & 0.7/49 \cr
3.0 - 5.5 & 1700 & $3.5^{+2.5}_{-1.0}$ & $0.6\pm0.1$ & 1.3/45 \cr
1.5 - 3.0 &700   & $4.0^{+5.0}_{-1.0}$ & $0.6\pm0.1$ & 1.1/32 \cr
centre - 1.0 & 270& $1.2^{+0.2}_{-0.1}$ & $0.9\pm0.4$ & 0.6/31\cr
centre - $1.0^a$&270& $-5.4^{+1.7}_{-2.2}$$^a$ &$6.5^{+3.5}_{-3.0}$&0.7/29\cr
SW: 1.3 - 7.0& 670 & $1.6^{+0.8}_{-0.2}$ &$0.8^{+0.3}_{-0.2}$ & 1.2/31\cr
SW: 1.0 - 1.3& 210 & $1.4^{+1.4}_{-0.2}$ &$0.4^{+0.2}_{-0.0}$ & 1.0/30\cr
\noalign{\hrule\medskip}\cr
}
\endtab

All the performed fits have in common, that the derived values for
$n_{\rm H}$ are 
within the errors in excellent agreement with the Galactic  value in the 
direction of A2634: $n_{\rm H}=0.52\times10^{21}~{\rm cm}^{-2}$ (Dickey
\& Lockman 1990).
Well constrained
temperatures are obtained 
in the centre of the cluster, $1.2^{+0.2}_{-0.1}$ keV (Fig. 7) and
in the outermost ring, $3.5^{+2.0}_{-1.0}$ keV (Fig. 8). 
The fit to the overall diffuse emission region is also well
constrained (Fig. 9), giving a cluster average 
temperature of $3.5^{+1.2}_{-0.7}$ keV. 
The spectra from the two inner rings are  statistically poorer. 
For these two rings, fixing $n_{\rm H}$ to the Galactic value
 sligthly modifies the $\chi^2$ of the fit, but hardly changes the
temperature. 

\begfig 0.0cm
\psfig{figure=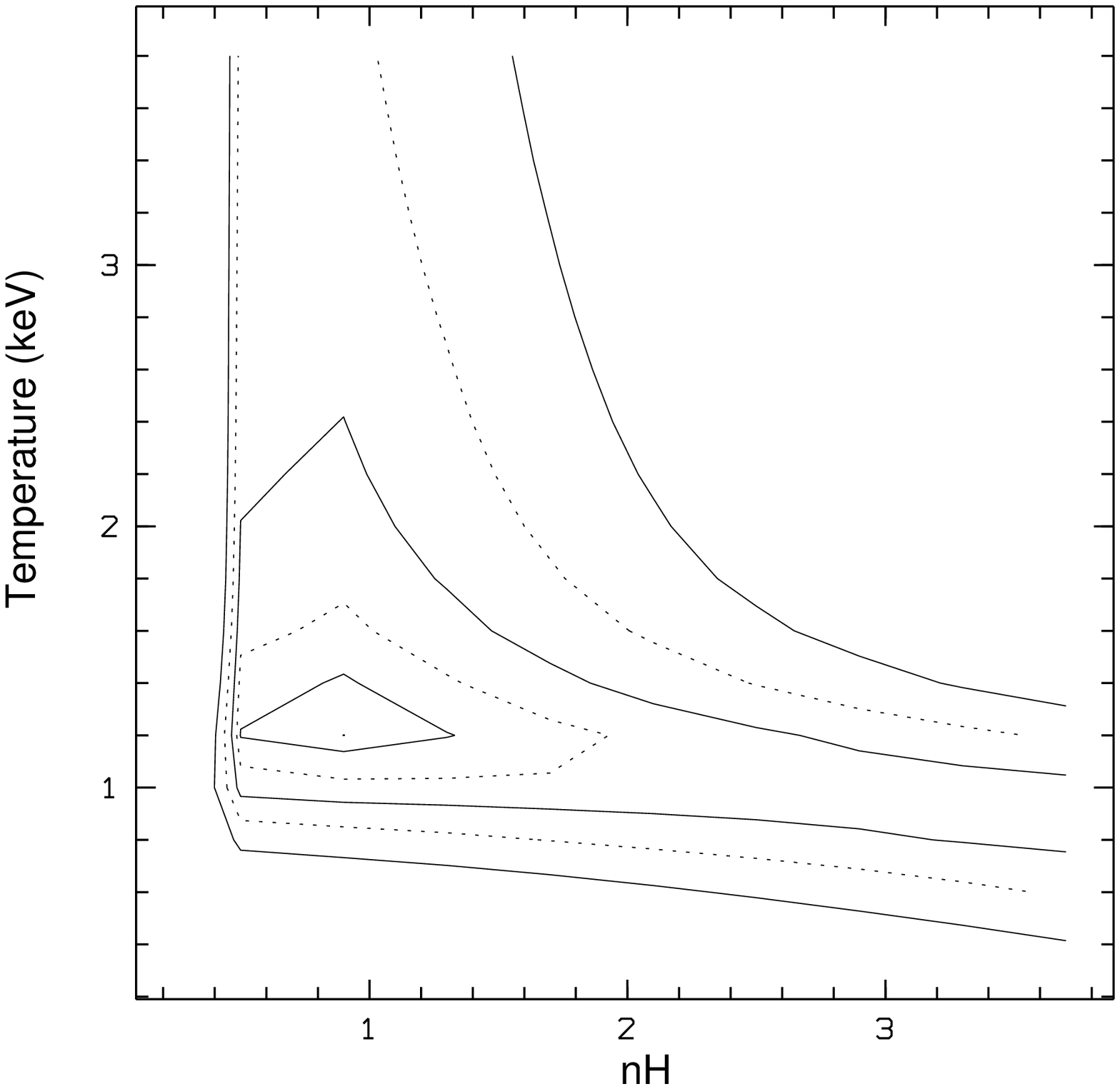,width=8.8cm,clip=} 
\figure{7}{Confidence contours for the fit parameters temperature 
and hydrogen column density ($n_{\rm H}$ in units of $10^{21}$ cm$^{-2}$)
for a Raymond-Smith-model fit to the central 1 arcminute region. 
The contours  span 1 to 5 $\sigma$.
}
\endfig

\begfig 0.0cm
\psfig{figure=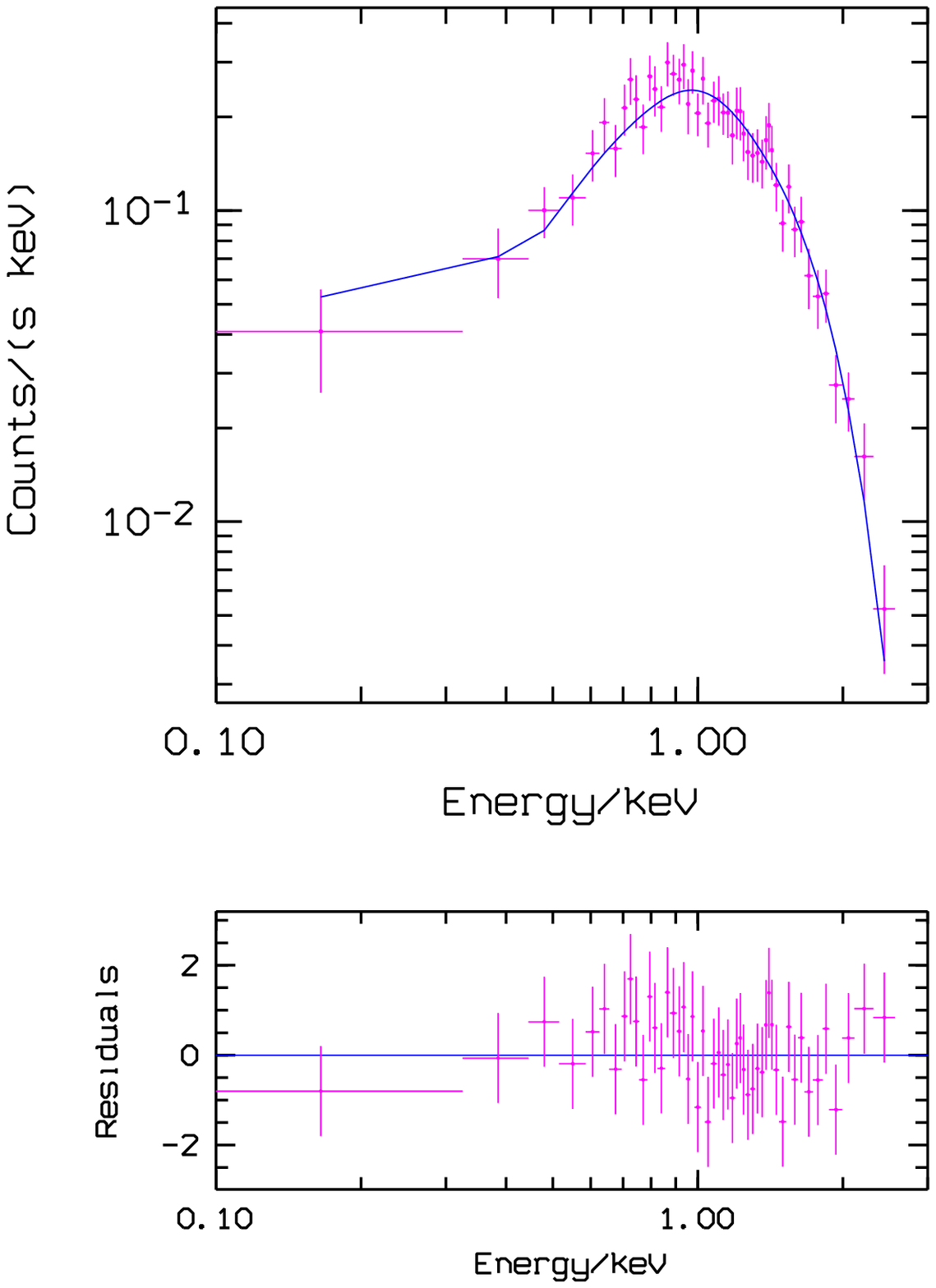,width=8.8cm,clip=} 
\figure{8}{PSPC spectrum of the cluster region between 3 
and 5.5 arcminutes from the centre. The solid line is a fit 
 with a Raymond-Smith model with fixed metallicity $m=0.35$ and fit
parameters as in Table 1.  The bottom panel shows the fit residuals in sigma 
units.}
\endfig

\begfig 0.0cm
\psfig{figure=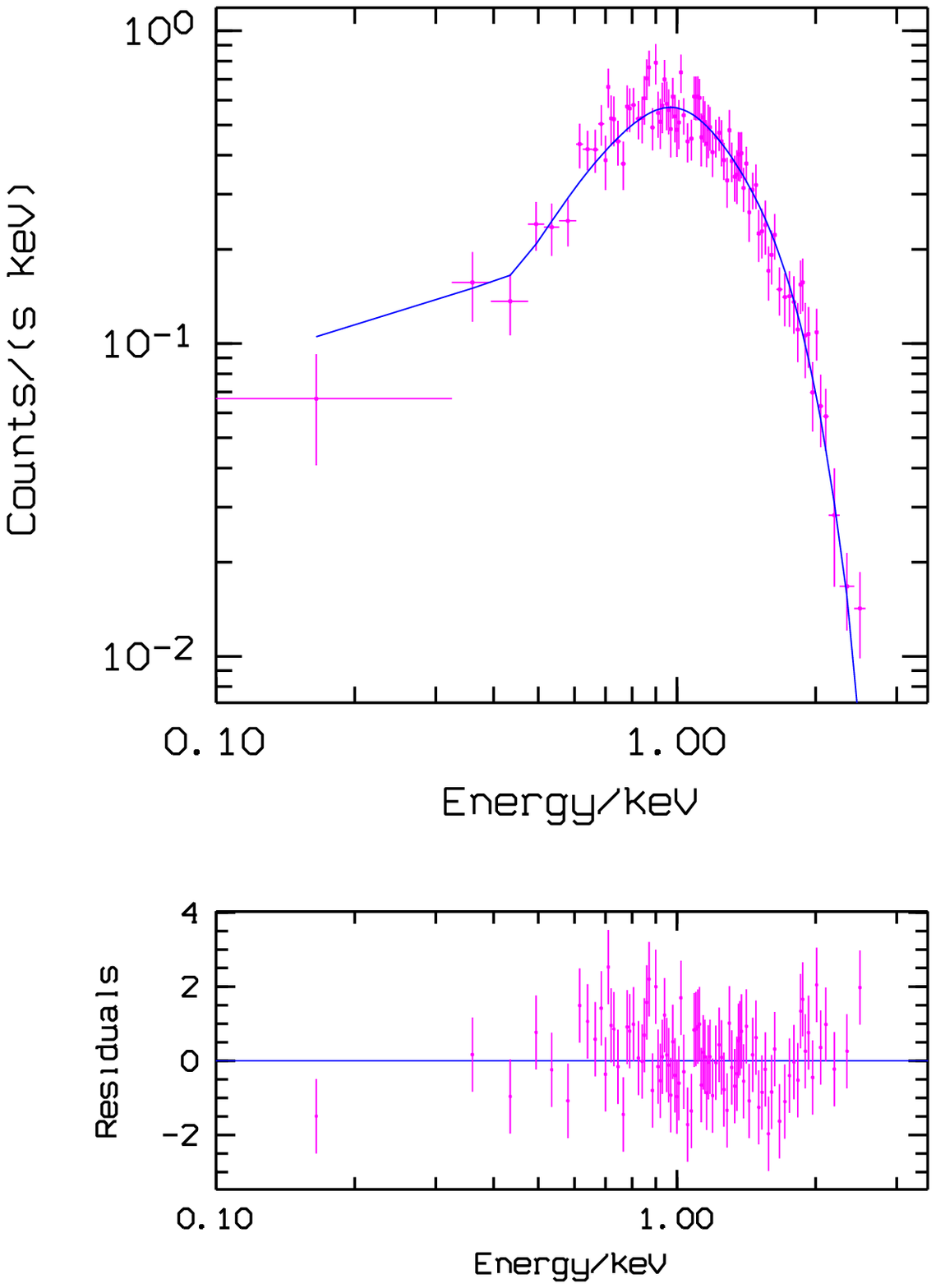,width=8.8cm,clip=} 
\figure{9}{PSPC spectrum of the cluster region  between 
 1.5 and 11 arcminutes from the centre. The solid line is a fit
 with a Raymond-Smith model with fixed metallicity $m=0.35$ and fit
parameters as in Table 1. The bottom panel shows the fit residuals in sigma 
units.}
\endfig

The  light profile shown in Fig. 5 indicates that the 
emission in the 
inner 1 arcminute of the cluster  is unresolved. It is dominated  by
strongly  peaked emission at 
the centre  compatible with the ROSAT/PSPC point-spread-function, 
the contribution of the extended 
gas emission  
being no larger than  1\%.
Since there is a radio galaxy in the centre of the cluster,
 one might expect most of the central emission to be of non-thermal
nature. For testing this hypothesis we fit
the spectrum from the central 1 arcminute region separately
with a power-law model and a thermal model. 
We find that the power-law 
model leads to extreme spectral indices and to an $n_{\rm H}$ being
more than an
order of magnitude higher than the Galactic value for a 
statistically acceptable fit (Fig. 10, Table 1). Therefore we consider this
model to be unrealistic. On the other hand 
a Raymond-Smith  model provides a good representation of
the central PSPC spectrum, leading to a $n_{\rm H}$ in fair agreement   with
the corresponding Galactic value (Fig. 7, Table 1). Thus the
{\it dominant} emission  appears to be of  thermal origin. 

\begfig 0.0cm
\psfig{figure=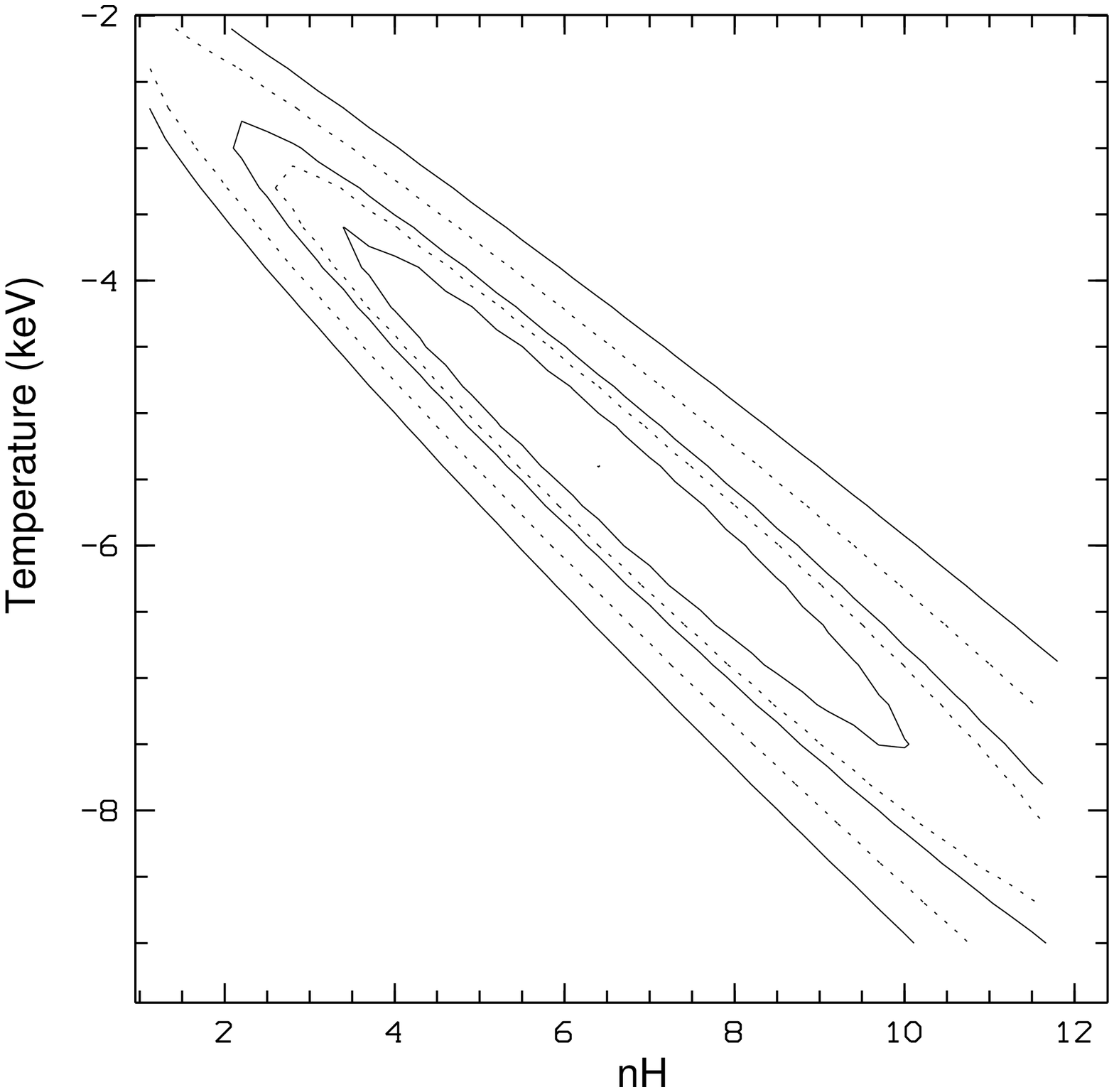,width=8.8cm,clip=} 
\figure{10}{Confidence contours for the fit parameters power-law index
and hydrogen column density ($n_{\rm H}$ in units of $10^{21}$ cm$^{-2}$)
derived from a power-law fit to the central 1 arcminute region of A2634.
The contours  span 1 to 5 $\sigma$.
}
\endfig

In addition, separate fits are attempted over several regions in the 
south-west region of the cluster. A first  fit is applied to the region 
between 1.3 and 7 arcminutes from the centre comprising angles between
  $180^\circ - 245^\circ$ (north over east). The central emission and the two 
blobs of bright  
emission seen  close to the centre are excluded. 
 The total number of counts, $\sim 670$ counts, slightly exceeds that
from other equivalent quadrants in the cluster, which have of the order of 400
counts on average. 
A temperature of
$1.6^{+0.8}_{-0.2}$ keV is found, which virtually fits  in between the
slighter lower temperature in the
central region  and the  higher one in the  outer regions of the 
cluster.

A second separate fit is attempted for the region comprising the two blobs of 
brighter emission seen  immediately close to the centre  of the cluster in 
the south-westerly direction.
The accumulated spectrum includes the region between  1 and 1.3 
arcminutes from the centre 
and  is defined  by the same angles as above. Although the accumulated 
 number of counts 
is relatively low,  $\sim$ 200, both temperature and  $n_{\rm H}$ are 
constrained in the fit. The temperature of $\sim$ 1.4 keV is  
just  following the general trend of decreasing temperatures towards the 
centre of the cluster.

Summarizing, a consistent decreasing temperature gradient towards the centre of
the cluster is measured in A2634. This is  supported by the
two fairly well constrained temperatures of about 3 and 1.2 keV 
derived for the outermost   and for the
inner 1 arcminute region of the cluster, respectively. The south-west
region of excess emission 
shows  marginal evidence for  lower temperatures below  3 keV.
A recent analysis of ASCA data of A2634 (Fukazawa, private
communication) is also consistent with a  
decreasing temperature towards the centre 
and with the south-west region   being of lower temperature.



%

\titlea{Luminosity}

The cluster emission can be traced out to a radius of about 30 arcminutes
(1.5 Mpc). Within this radius  a count rate of 1.1
counts/s (only source counts; excluding the emission of CL-37) is obtained. 
Assuming an average cluster temperature of $3.5^{+1.2}_{-0.7}$ keV (as
derived in Sect. 4, see Table 1), a
Galactic hydrogen column 
density of $0.52\times10^{21}$cm$^{-2}$  
and an average cluster
metallicity of 0.35 solar, that count rate 
corresponds to a luminosity of $7.9\pm0.1\times 10^{43}$ erg/s in the
ROSAT band (0.1 - 2.4 keV) and  a bolometric luminosity of
$1.4\pm0.3\times 10^{44}$ erg/s. The errors only refer to the
uncertainty in the temperature.

The luminosity in the central 1 arcminute inner  region is
$2.2\times10^{42}$erg/s (0.1-2.4 keV). This has been derived using the
parameter values listed in Table 1. The contribution from the 
extended
gas emission is 
inferred, following  the spatial analysis of Sect. 3 (Fig. 5), to be 
less than 1\%.

\titlea{Mass determination}

The fit parameters of the $\beta$ model: $\Sigma_0=3.4\times 10^{-3}$
counts/s/arcmin$^2$, $r_c= 9.8$arcmin and $\beta = 0.79$,
are used to make a deprojection of the 2D image to get the three
dimensional density distribution. 
In doing so, it is assumed  that the clumpiness of the gas and local
temperature variations do not affect the mass
estimate grossly. The corresponding  gas mass profile is shown in  Fig. 11.
Within a radius of 1.5 Mpc, the gas mass amounts to
$5.1\times 10^{13}\msol$. 

With the additional assumption of hydrostatic 
equilibrium, the integrated {\it total} mass is  calculated from the equation
$$
M(r) = {-kr\over \mu m_p G} T \left({ d \ln \rho \over d \ln r }+
                                    { d \ln T    \over d \ln r }\right),
   \eqno(2)
$$
where $\rho$ and $T$ are the density and the temperature of the
intra-cluster gas, and $r$, $k$, $\mu$, $m_p$, and $G$ are the 
radius, the Boltzmann constant,
the molecular weight, the proton mass, and
the gravitational constant, respectively. 
The assumption of hydrostatic equilibrium is well justified even when
assuming that the excess emission in the south-west region  corresponds
 to an
infalling subcluster. This excess luminosity of this region is less than 2\% of
the total luminosity of the cluster. Converting this luminosity fraction to a
mass fraction we find that the subcluster represents  only 10\% of the
total cluster mass 
which is too  small to introduce  significant deviations from an
 equilibrium configuration.

In the central 1 arcminute, a mass determination is hardly possible,
because the most important factor in Eq. 2 -- the density gradient --
is dominated by the point spread function of the PSPC. Therefore, this
region of the cluster is
excluded when deriving the total mass. The rest of the cluster is
considered to be  
isothermal with  an  average temperature of 
$3.5^{+1.2}_{-0.7}$ keV (Table 1). The
profile of the integrated mass is
shown in Fig. 11. As we cannot determine the temperature beyond 
550 kpc, we assume that the temperature in the outer parts 
is the same as between 75 and 550 kpc.
If using the $\beta$ fit parameters that are derived from  
the north-east part
of the cluster, i.e. the region less affected by substructure (Sect. 3), the 
corresponding   mass profile is essentially the same.

At a radius of 1.5 Mpc the derived  total mass is 
$4.1^{+2.6}_{-1.8}\times 10^{14}\msol$. The errors only reflect  the 
uncertainty
in the cluster temperature (including possible temperature
gradients within the error range). 
The systematic errors are probably much smaller
(Schindler 1996a; Evrard et al. 1996). 
For comparative purposes, the total mass derived from the X-ray
observation is in agreement with the
viral mass, 
$5.2\times10^{14}\msol$,   derived by  Scodeggio et al. (1995) for the same 
cluster radius. 

 The gas mass
fraction at 1.5 Mpc radius 
is found to be 12$^{+10}_{-5}$\% and  increases very slightly with radius.

\begfig 0.0cm
\psfig{figure=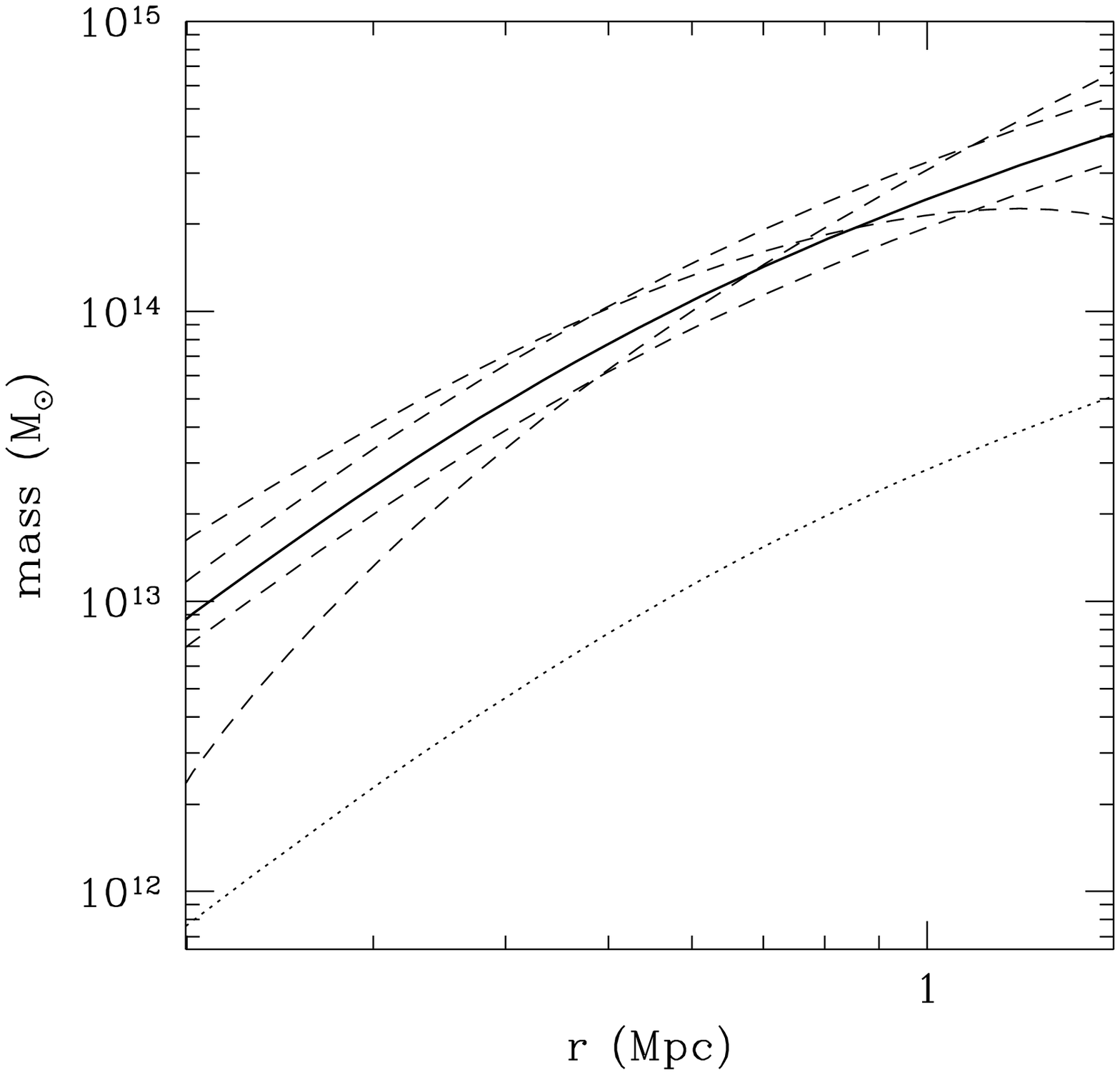,width=8.8cm,clip=} 
\figure{11}{Profile of the integrated total mass (full line) 
between 0.2 and 1.5 Mpc with
errors coming from
the uncertainty in the overall temperature and possible temperature
gradients within the error range (dashed lines). 
The dotted line
shows the profile of the integrated gas mass. 
}
\endfig

\titlea{Central emission}

As shown in Sect. 4, the  centrally peaked
emission in A2634 is well described by a single, {\it thermal} spectral
 component. Though  non-thermal emission due to the  central radio source 
would also be  expected, the  thermal component appears, however, 
 to be the dominant component of the emission.
 
 The integrated 0.1-2.4 keV luminosity 
from  the inner 1 arcmin, $L_X = 2.2\times 10^{42}$erg s$^{-1}$,
 places A2634 within the lower 
power range of X-ray luminosities measured in the central region of  clusters 
dominated by FRI radio sources (Prieto 1996). In general, the  
X-ray  luminosities in those cases span a range of $10^{42} - 
10^{44}$erg s$^{-1}$, and mostly reflect the cluster  emission. 

The  central $n_{\rm H}$ derived 
from the spectral fit  falls within the average range 
of values found in our Galaxy. Thus it is unlikely that high
absorption is the cause for the 
AGN source to be undetected. An HST/WFPC2 image of 3C465 in the filter
F702W reveals clearly a 
thick ring of dust surrounding the galaxy bulge (Fig. 2).
  The central source may be partially
obscured by this material but not completely as the plane of the
ring is almost face-on with respect to the line of sight. 
All facts together suggest that most probably the AGN activity in 3C465 is 
presently very weak.
 
 3C465 is  a cD galaxy. This property together with the measured 
temperature drop in the
centre of the cluster are suggestive of a cooling flow scenario. As it is 
shown below, parameters derived from   
 the present X-ray data are
compatible with a weak cooling flow.


The  central electron density of the cluster is determined using 
 a {\it composite} 
 $\beta$ model. The  $\beta$
model used in Sect. 3  accounts well for the emission distribution
beyond a radius of 2 arcminutes from the centre (see Fig. 5); yet, if only
this model would be used to 
derive the  central 
density, the density would be underestimated by more than an order of
magnitude. 
Thus, to account properly for the 
central emission, a second $\beta$ model is fit to the
residual image resulting
after subtracting the  $\beta$ model of Sect. 3. This second $\beta$ 
model fits the residual image up to a radius of 37 arcseconds from  
the centre.

An estimate of the cooling flow parameters are derived from 
the sum of both $\beta$ models. The central electron density is found
to be  
0.016cm$^{-3}$, which yields a central cooling time of about $3\times10^9$
yr. This time is considerably smaller than the Hubble time, thus an
equilibrium situation could  have established. With the usual
assumptions (i.e., a cooling time smaller than $10^{10}$ yr) a cooling
flow radius of about 25 arcseconds ($\sim$ 20 kpc) is derived.
 Using  the approximation:

$$L_{\rm cool} = {5 \over 2}{\dot{M} \over \mu m} kT,
   \eqno(3)
$$
the derived mass accretion rate  within that radius is about
1 $\msol$/yr. Unfortunately, the central emission is not resolved (see
Fig. 5), therefore we cannot see the true extent of the cooling
flow. Thus the derived radius of 20 kpc is actually an upper limit.

\titlea{Comparison with optical data}

 Evidence for cluster substructure has been  found in a substantial
number of
EINSTEIN images (e.g. Forman \& Jones 1982; Mohr et al. 1993) and presently,
better resolved,  in   ROSAT images 
 (e.g. A2256, Briel et al. 1991;  Virgo, B\"ohringer et al. 1994; 
A3528, Schindler 1996b; A3627, B\"ohringer et al. 1996; CL0939+4713,
Schindler \& Wambsganss 1996; CL0016+16, Neumann \& B\"ohringer 1997). 
Regions of excess X-ray emission in clusters are usually
interpreted as the signature of   a subcluster or  galaxy group being 
in the process of merging  
with  the main cluster.  As such, the south-west X-ray enhancement 
in A2634 may also be indicative of  subclustering.

Potential optical counterparts to the X-ray excess region are sought 
 on the basis of the  optical study of the   cluster members by Scodeggio et 
al. (1995).
Judging from the apparent spatial distribution of the cluster galaxies, no 
obvious  correspondence with either individual
 members or groups of them is found.
Indeed, Scodeggio et al.  remarked on   the lack of  significant clumpiness in 
both the  kinematic 
 and the spatial  distribution of the  galaxy members. They found, however, a 
clear
 morphological segregation
 between the early- and the late-type galaxy population: E and S0 are
spread all over the cluster;  S and Irr are almost absent in the central 
region of
the cluster. Since the region of  X-ray excess 
 arises in the inner  ($\sim 6$ arcmin) region of the cluster, it  
may only be related, if anything, to  the early-type population members.
Focusing  on the central
 20 arcminutes of the cluster (cf. Fig. 14 in Scodeggio et al. 
1995), a certain level of alignment of the  cluster galaxies 
in the direction close to  the major axis of  
3C465  is apparent. This direction is roughly coinciding with  that marked by 
the south-west excess region. While the combined
effect could just reflect an asymmetric central potential, we note that,
first, the galaxies concentration   is more accentuated
in the north-east side of the cluster rather than
 in south-west where
most of X-ray excess arises; second, the overall 
extended
emission appears compatible with spherical symmetry (see Sect.3).

If focusing  on  the kinematic of the early-type
population members,   their
spatial and kinematic distribution of velocities (Scodeggio et al. 
1995) are also found  compatible with those expected from a dynamically
relaxed system. 

Thus, on the basis of both kinematic and morphology of the cluster galaxies,
 the observed X-ray excess in A2634 is
unlikely to be associated with  a subcluster or  group falling into the
cluster.
As a final cross-check, the Palomar plates are examined 
for potential unrelated sources lying in  the  
direction of the cluster. Both red and blue plates are examined and no
association with any outstanding X-ray feature is found. Therefore,
 on the basis   of the available optical data,  the
X-ray excess emission in the south-west region of A2634 appears to be 
unrelated with any visible mass distribution.

\titlea{The X-ray / radio relationship}

Fig. 3 compares the X-ray and radio emission  in A2634. 
 Projection effects hamper a clear determination of the relative 
distribution of both
 emissions and thus, on their possible interaction. 
Nevertheless, Fig. 3 clearly shows the 3C465 radio jet  extending
 perpendicular to the direction of X-ray excess region. 
Also, the  wide angle tails  that characterize the radio structure 
bend along the same {\it projected} directions 
on the sky, thus, are virtually embracing the south-west X-ray excess region.

An alternative  way to probe the dynamical state of  the  system is to 
verify whether the gas and plasma pressures are in equilibrium. To this end, 
equipartition pressures in the tails, 
as derived by Eilek et al. (1984), are compared with  
thermal 
pressures   derived from the present ROSAT data. The latter 
are obtained under the assumption of
 spherical symmetry. The density values derived  from
 the two added $\beta$ models (Sect. 7) are adopted.

For those regions in the tails far  away from the centre, E4, F, A1 and
A2 (see Fig. 3; nomenclature by Riley \& Branson 1973), thermal and
equipartition pressures are found in balance, in agreement with what
was found by Eilek et al. (1984) on the basis of EINSTEIN IPC data.
However, for those regions closer to the centre, i.e., D and B1, which
are roughly at the position where respective tails begin to open,
thermal pressures are found a factor 2 - 5 smaller than the
equipartition pressures. For the intermediate regions no temperatures
are available (see Table 1); yet, when interpolating between the inner
and the outermost temperature an increasing discrepancy towards the
centre is found. While individual pressure values may be uncertain due to 
the number of
 involved assumptions, the tendency for an increasing pressure imbalance
towards the centre of the cluster should be regarded as a more solid
result, and it would imply  that the system is not yet in an equilibrium 
configuration.

\titlea{Discussion and Conclusions}

Two particular features characterize the  X-ray image of A2634: strongly
peaked emission at the centre and a region of enhanced   emission in the
south-west area of the cluster. In addition, the system is embedded in 
a spherical  gas region  which can be
traced up to $\sim$1.5 Mpc from the cluster centre.

Within the  spatial and spectral limitations of the PSPC data, 
  a cooling flow scenario arises as the best interpretation for the  central 
emission in A2634. There are several arguments supporting  this
scenario: the thermal nature of 
the central spectrum, the presence of a  cD
galaxy and the  temperature drop at the centre (the central temperature
 is found to be 1.2 keV while the average intra-cluster gas is about
3.5 keV). The temperature drop is also confirmed by 
ASCA (Fukazawa, private communication).

Within the cooling flow radius of 20 kpc we derive a
relatively small mass accretion rate of about 1 $\msol$/yr.
The   mass accretion rate
falls, nevertheless, within the   range of values derived from  the  
correlation between  mass accretion rates, luminosity and temperature in
cooling flow clusters (Fabian et al. 1994).
 For comparative purposes,  the prototype  cooling flow in Virgo (M87)
presents  a cooling
flow radius of 70 kpc (also for a cooling time smaller than $10^{10}$ yr) and 
a mass accretion rate of about 10 $\msol$/yr  (Stewart et
al. 1984).  The situation of A2634 seems closer to  Fornax, where a
mass accretion  
rate of 2 $\msol$/yr within a
radius of 40 kpc was derived (Rangarajan et al. 1995). The cooling
flows in Virgo and Fornax
could, however, be resolved as  they are much
closer than A2634.

The expected AGN contribution  due to the central source 3C465
is not readily detected in the PSPC band. The 0.1-2.4 keV  central 
spectrum appears instead  
 dominated by a thermal component,  moderately absorbed by a $n_{\rm H}$ in
close agreement with the  Galactic value.
The fact that no  excess absorption is found towards the cluster centre 
lead us to the conclusion that the AGN source rather than be
obscured is  in a quiescent period. 

The presence of enhanced X-ray emission in the south-west  region of the
cluster motivates the  search for  optical counterparts that could  
be related to  a potential subcluster or group  in interaction with
A2634. Previously, Pinkney et al. (1993) and Burns et al. (1993) already 
suggested a merging 
scenario for A2634. However, on of the basis of both observational 
and theoretical considerations we find the  merging scenario unlikely. The 
 X-ray excess region shows no correspondence with   the 
 spatial distribution of the cluster members, nor with their 
corresponding spatial distribution of velocities. A comparison with
hydrodynamic simulations  
of subcluster collisions (Schindler \& M\"uller 1993)  shows that in  
the case of a {\it post-merging phase}, the intracluster 
 gas  is squeezed out
by the merger   to show up as hot bulges in
directions perpendicular to the collision axis. 
However,   the  region of enhanced X-ray emission in A2634
shows a well defined elongated distribution which
is likely defining, if anything, 
the direction of the potential encounter. On the other hand,  in  a
{\it pre-merging phase}, the simulations predict the formation of a 
heated region between the two parties as soon as they get closer 
than about 1-2 Mpc. 
Such a feature is not seen in the temperature
map of the cluster. Instead,  the temperature measurements in the 
south-west region are
consistent with a progressive decay towards the 
centre of the cluster. This temperature drop in south-westerly
direction is also confirmed by ASCA data (Fukazawa, private
communication). 

 A2634 can be classified as a poor cluster.
It has a low  X-ray luminosity,
$1.4\pm0.3\times 10^{44}$erg/s (bolometric) 
if compared with other clusters of similar optical richness (class I).
The inferred mass,   $4.1^{+2.6}_{-1.8}\times 10^{14}\msol$,
 is at the lower limit of the typical
mass range in clusters, $5-50\times10^{14}\msol$, as well 
as the gas mass fraction, 
 about 12\%,  compared with   typical  values  
in clusters, 10-30\% (B\"ohringer 1995).

On the other hand, the basic  cluster parameters   are consistent with each 
other and with the general  relations found in clusters.
Relative to the bolometric luminosity the  average cluster temperature
falls into the
expected range  (e.g. Edge \& Stewart 1991a; David et al. 1993; White 1996).
The velocity
dispersion of 661 km/s  (Scodeggio et al. 1995)
also lies within the $\sigma-T$ relation (Edge \& Stewart 1991b;
Lubin \& Bahcall 1993; Bird et al. 1995). 

A2634 has many properties in common with the Virgo cluster: It has
about the same luminosity and is only slightly more 
massive (B\"ohringer 1995). Virgo is a prototype cooling flow, and 
also shows X-ray substructure   (B\"ohringer et al. 1994).
The associated
radio emission in M87 is, however, correlated in position with the
X-ray emission (B\"ohringer et al. 1995), 
while in A2634 the radio structure opens its way along
directions perpendicular to the region of enhanced X-ray emission.
This anti-correlation resembles the situation in the Perseus cluster,
where the X-ray emission shows minima at the positions of the radio
lobes (B\"ohringer et al. 1993) suggesting that the pressure of the 
relativistic particles has displaced the thermal gas. 

The radio emission associated with 3C465 shows a well defined configuration.
The 3C465 radio jet is seen extending perpendicular to the direction
of enhanced X-ray emission and both radio tails appear to embrace
virtually the south-west X-ray excess region. Pressure equilibrium
between the relativistic plasma and the intra-cluster medium
is found at the outer parts of the radio
tails. However, an increasing imbalance is measured towards the inner regions 
of the cluster,  the plasma pressure being found larger than that
of the gas. This overpressure is indicative that the system may not have 
reached  an equilibrium configuration yet.  
No visible mass is seen associated with the south-west excess region, so it is 
plausible that part of the X-ray substructure 
seen  in that region just reflects an
inhomogeneous density medium, perhaps being produced by the displacement of 
the gas  by the plasma flow.
 The decreasing temperature 
gradient measured in this region would then be  consistent with the 
formation of cooler gas  
condensations  in equilibrium with the surrounding intra-cluster gas.

\begtabfull
\tabcap{2}{Summary of the X-ray properties of A2634}
\halign{#\quad\hfil&
#\quad\hfil \cr
\noalign{\hrule\medskip}\cr
$L_X$(0.1-2.4keV)      & $7.9\pm0.1\times 10^{43}$erg/s\cr
$L_X$(bol)             & $1.4\pm0.3\times 10^{44}$erg/s\cr
count rate(0.1-2.4keV)  & $1.1$counts/s \cr
$r_c$                  & $9.8^{+1.3}_{-1.1}$arcmin ($490^{+70}_{-50}$kpc)\cr
$\beta$                & $0.79^{+0.10}_{-0.07}$      \cr
$M_{gas}$($r<1.5$Mpc)    & 0.51$\times 10^{14}\msol$\cr
$M_{tot}$($r<1.5$Mpc)    & $4.1^{+2.6}_{-1.8}\times 10^{14}\msol$\cr
gas mass fraction      & 12$^{+10}_{-5}$\% \cr         
central cooling time   & $3\times10^9$yr  \cr
mass accretion rate    & $\sim 1 \msol$/yr      \cr
of possible cooling flow &  \cr
\noalign{\hrule\medskip}\cr
}
\endtab

\acknow{}
It is a pleasure to thank Hans B\"ohringer
and Doris Neumann for helpful discussions. 
S.S. acknowledges 
financial support by the Verbundforschung.

\begref{References}

\ref  Arnaud M., Rothenflug R., Boulade O., Vigroux L., Vangioni-Flam
      E., 1992, A\&A 254, 49 

\ref  Bender R., M\"ollenhoff C., 1987, A\&A 174, 63 
\ref  Bird C.M., Mushotzky R.F., Metzler C.A., 1995, ApJ 453, 40
\ref  B\"ohringer H., 1995, in: Proceedings of the 17$^{th}$
        Texas Symposium on Relativistic Astrophysics and Cosmology, 
	B\"ohringer H., Morfill G.E., Tr\"umper J.E. (eds.), 
	New York Academy of Sciences, New York, p. 67
\ref  B\"ohringer H., Voges W., Fabian A.C., Edge A.C., Neumann D.M.,
      1993, MNRAS 264, L25
\ref  B\"ohringer H., Briel U.G., Schwarz R.A., Voges W., Hartner G.,
        Tr\"umper J., 1994, Nat 368, 828
\ref  B\"ohringer H., Nulsen P.E.J., Braun R., Fabian A.C., 1995,
        MNRAS 274, L67
\ref  B\"ohringer H., Neumann D.M., Schindler S., Kraan-Korteweg R., 1996,
        ApJ 467, 168
\ref  Briel U.G., Henry J.P., Schwarz R.A.,
       B\"ohringer H., Ebeling H., Edge A.C.,
       Hartner G.D., Schindler S., Tr\"umper J., Voges W., 1991, A\&A
       246, L10
\ref  Burns J.O., Rhee G., Roettiger K., Owen F.N., 1993, in: ASP
        Conference Series, Vol. 51, Observational Cosmology,
        Chincarini G., Iovino A.,
        Maccacaro T., Maccagni D. (eds.), ASP, San Francisco, p. 407
\ref  Cavaliere A., Fusco-Femiano R., 1976, A\&A 49, 137
\ref David L.P., Slyz A., Jones C., Forman W., Vrtilek S.D.,
        Arnaud K.A., 1993, ApJ 412, 479
\ref de Juan L., Colina L., P\'erez-Fournon I., 1994, ApJS 91, 507

\ref  Dickey J.M., Lockman F.J., 1990, ARA\&A 28, 215

\ref Edge A.C., Stewart G.C., 1991a, MNRAS 252, 414
\ref Edge A.C., Stewart G.C., 1991b, MNRAS 252, 428
\ref Eilek J.A., Burns J. O., O'Dea C.,  Owen F.N., 1984, ApJ 278, 37
\ref Evrard A.E., Metzler C.A., Navarro J.F., 1996, ApJ 469, 494
\ref Fabian A.C., Crawford C.S., Edge A.C., Mushotzky R.F., 1994, 
        MNRAS 267, 779 
\ref  Forman W., Jones C., 1982, ARA\&A 20, 547
\ref  Jones C., Forman W., 1984, AJ 276, 38

\ref  Laing R.A, Riley J.M., Longair M.S., 1983, MNRAS 204, 151 
\ref  Lubin M.L., Bahcall N.A., 1993, ApJ 415, L17
\ref  Mohr J.J., Fabricant D.G., Geller M.J., 1993, ApJ 413,492 
\ref  Neumann D.M., B\"ohringer H., 1997, MNRAS 289, 123
\ref  Pinkney J., Rhee G., Burns J.O., Hill J.M., Oegerle W., Batuski
      D., Hintzen P, 1993, ApJ 416, 36
\ref  Prieto M.A., 1996, NRAS 282 
\ref  Rangarajan F.V.N., Fabian A.C., Forman W.R., Jones C., 1995,
      MNRAS 272, 665 
\ref  Raymond J.C., Smith B.W., 1977, ApJS 35, 419

\ref  Riley J.M., Branson N.J.B.A., 1973, MNRAS 164, 271
\ref  Sakelliou I., Merrifield M.R., 1997, MNRAS, submitted
\ref  Schindler S., 1996a, A\&A 305, 756
\ref  Schindler S., 1996b, MNRAS 280, 309
\ref  Schindler S., M\"uller E., 1993, A\&A 272, 137 
\ref  Schindler S., Wambsganss J., 1996, A\&A 313, 113
\ref  Scodeggio M., Solanes J.M., Giovanelli R., Haynes M.P., 1995,
      ApJ 444, 41
\ref  Stewart G.C., Canizares C.R., Fabian A.C., Nulsen P.E.J., 1984,
      ApJ 278, 536
\ref  Tsuru T., Koyama K., Hughes J.P., Arimoto N., Kii T., Hattori M.,
      1996, in: UV and X-Ray Spectroscopy of Astrophysical and Laboratory
      Plasmas, Yamashita K., Watanabe T. (eds.), Universal Academic
      Press, Tokyo,  p. 375
\ref Venturi T., Castaldini C., Cotton W.D., Feretti L., Giovannini
     G., Lara L., Marcaide J.M., Wehrle A.E., 1995, ApJ 454, 735
\ref  White D.A., 1996, in: R\"ontgenstrahlung from the Universe,
      Zimmermann H.U., Tr\"umper J.E., Yorke H., MPE report 263, p. 621
\ref  Yamashita K., 1994, in: Clusters of Galaxies, Durret F., Mazure
      A., Tr\^an Thanh V\^an J. (eds.), Edition Frontieres, Gif-sur-Yvette,
      p. 153

\endref
\bye